\documentclass[twocolumn,aps,superscriptaddress]{revtex4-1}
\usepackage[utf8x]{inputenc}
\usepackage{ucs}
\usepackage{amsmath}
\usepackage{amsfonts}
\usepackage{amssymb}
\usepackage{makeidx}
\usepackage{graphicx}
\usepackage{xcolor}
\usepackage{listings}
\usepackage{subfig}

\begin{document}
\author{M. H. Jensen}
\email{mihaje@ruc.dk}
\affiliation{DNRF Centre Glass and Time, IMFUFA, Department of
  Sciences, Roskilde University, Postbox 260, DK-4000 Roskilde,
  Denmark} 
\affiliation{Laboratoire Léon Brillouin, CNRS CEA -UMR 12, DSM IRAMIS
  LLB CEA Saclay, 91191 Gif-sur-Yvette Cedex, France}

\author{C. Alba-Simionesco} 
\affiliation{Laboratoire Léon Brillouin, CNRS CEA -UMR 12, DSM IRAMIS
  LLB CEA Saclay, 91191 Gif-sur-Yvette Cedex, France}

\author{K. Niss}
\affiliation{DNRF Centre Glass and Time, IMFUFA, Department of
  Sciences, Roskilde University, Postbox 260, DK-4000 Roskilde,
  Denmark} 

\author{T. Hecksher}
\affiliation{DNRF Centre Glass and Time, IMFUFA, Department of
  Sciences, Roskilde University, Postbox 260, DK-4000 Roskilde,
  Denmark} 

\title{A systematic study of the isothermal crystallization of the
  mono-alcohol n-butanol monitored by dielectric spectroscopy}
\date{\today}

\begin{abstract}
  Isothermal crystallization of the mono-hydroxyl alcohol n-butanol
  was studied with dielectric spectroscopy in real time. The
  crystallization was carried out using two different sample cells at
  15 temperatures between 120~K and 134~K. For all temperatures, a
  shift in relaxation times to shorter times was observed during the
  crystallization process, which is characterized by a drop in
  relaxation strength. The two different sample environments induced
  quite different crystallization behaviors, consistent and
  reproducible over all studied temperatures.  An explanation for the
  difference was proposed on the background of an Avrami and a
  Maxwell-Wagner analysis. Both types analysis suggest that the
  morphology of the crystal growth changes at a point during the
  crystallization. The differences between the cells can be explained
  by this transition taking place at different times for the two
  cells.
\end{abstract}

\maketitle

All liquids can be supercooled \cite{Tammann1925, Kauzmann1948,
  Debenedetti1996}. In fact, crystallization rarely takes place
exactly at the melting temperature upon cooling, because the crystal
nuclei formed dissolve before they grow to a stable size
\cite{Becker1935}. Some liquids, like water, crystallize readily at
moderate supercooling and need fast quenching below the melting
temperature in order to avoid crystallization. Others, like the
prototype glass-former glycerol, supercool easily and require a
careful protocol to crystallize \cite{Yuan2012}. But the true
thermodynamic equilibrium state for all supercooled liquids and
glasses is unarguably the crystal, and thus crystallization is their
eventual inevitable fate.

For many applications the life-time of the glassy or meta-stable
liquid state is a key issue and the ability to predict and control
crystallization properties is desired. But understanding the
crystallization process is also interesting from a fundamental point
of view. Crystallization studies are however difficult to carry out in
a controlled and reproducible manner, because many factors influence
the initiation and course of crystallization, such as sample
preparation, thermal history, presence of impurities, container
geometry, etc \citep{Rabesiaka1961,Yuan2012,Napolitano:2006bd}.

Dielectric spectroscopy is a convenient and fairly common probe for
studying crystallization kinetics in real time, see e.g. Refs.
\citep{MassalskaArodz:1999fta, Alie:2003ud, Adrjanowicz2010,
  Dantuluri:2011gu, Kothari:2014go, Sibik:2014dl}. The sign of
crystallization is a decrease in intensity of the signal, and the
crystal fraction/concentration of the sample is often obtained by
assuming that the relaxation strength is proportional to the volume
fraction of liquid in the sample \cite{Ezquerra1994, Viciosa:2009hr,
  Adrjanowicz2014}. But the microscopic interpretation of dielectric
spectra is not straight-forward \cite{Kremer2003a}, especially when
studying a heterogeneous mixture.

We present a thorough dielectric study of the temperature dependence
of the crystallization process in n-butanol. Supercooled n-butanol has
an intense low-frequency dielectric signal -- the so-called
Debye-process -- characteristic of many monohydroxyl alcohols
\cite{Bohmer2014}. It also exhibits a slow crystallization process
upon reheating after a rapid quench below $T_g$ \citep{Shmytko:2010hp,
  Derollez:2013dy}. At 10~K above the glass transition temperature the
crystallization can take several days to finish, and at temperatures
close to the glass-transition, the crystallization process stops
before the sample is fully crystallized \cite{Hedoux:2013ep}. These
two properties, a large dielectric signal and slow crystallization,
makes n-butanol an ideal candidate for monitoring isothermal
crystallization in real time by dielectric
spectroscopy. Crystallization of n-butanol has previously been studied
with x-ray diffraction \citep{Shmytko:2010hp,Derollez:2013dy},
infrared spectroscopy \citep{Wypych:2007kl,Hedoux:2013ep},
calorimetric methods \citep{Hassaine:2011gx}, and phase contrast
microscopy \citep{Kurita:2005jr}. However, no systematic study of the
temperature dependence and reproducibility of the crystallization
kinetics has been reported. We used two different dielectric measuring
cells with different geometry and different electrode material in an
attempt to disentangle effects that are intrinsic to the sample and
effects that are due to macroscopic boundary conditions.

\section{Experiment and materials}\label{sec:exp}

\begin{figure}[ht!]
\includegraphics{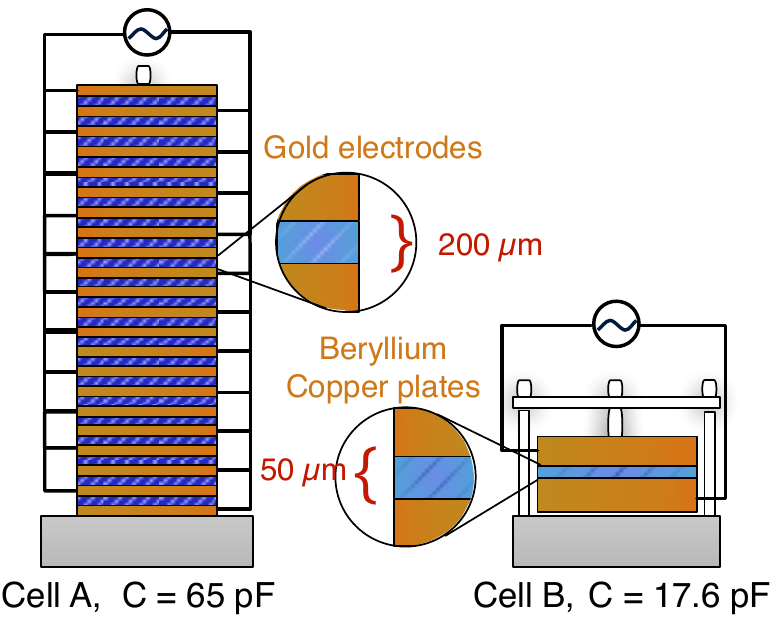}
\caption{\label{fig:cells}Schematic drawings of the two
  cells used for the crystallization studies. Cell A is a multilayer
  capacitor (variable capacitor type used in old fashioned radios)
  with a geometric capacitance of 65~pF. Cell B is a two-plate
  capacitor with beryllium copper electrodes and 50~$\mu$m sapphire
  spacers. Cell B has a geometric capacitance of 17.6~pF.}
\end{figure}

All measurements were carried out in the same experimental set-up
(described in detail in Ref. \onlinecite{Igarashi2008a}), including a
custom-built nitrogen cryostat capable of keeping the temperature
stable within 50~mK over weeks. Two different sample cells were used.
Cell A is a 22-layered gold-plated parallel plate capacitor with 0.2
mm between each set of plates and a geometrical capacitance of 64
pF. Each plate is a semi-circle which can be rotated to overlap each
other (identical to the capacitors used in old radios). Cell B is a
parallel plate capacitor with circular beryllium-copper plates
separated by 50~$\mu$m sapphire spacers. The cells are sketched in
Fig.\ \ref{fig:cells}.

The sample n-butanol ($T_m = 183 K$, $T_g \approx 110 K$) was
purchased from Sigma Aldrich's at $>99.9\%$ purity and used without
further purification.

The same protocol was followed for each crystallization measurement; a
new sample was quenched to 85~K (roughly 25~K below $T_g$) and kept at
this temperature for (at least) 120 minutes, then heated to the target
temperature where the crystallization process was followed. The
heating took less than five minutes. Frequency scans were made
continuously as soon as heating from 85~K initiated and until no
further changes in the spectrum occurred. The frequency range of the
scans was adjusted for each temperature to keep the scan as short --
and thereby as fast -- as possible, while still keeping both
$\epsilon_\infty$ and $\epsilon_s$ in the frequency window. The sample
cells were emptied and cleaned between each measurement.

Isothermal crystallization was followed at 15 different temperatures
between 134 and 120 K. The exact temperatures in Kelvin are: 134, 133,
132, 131, 130, 129, 128.5, 127.5, 126, 125, 124, 123, 122.5, 121, 120.
The 133~K measurement has only been done with cell A, while the 132~K
measurement has only been done with cell B. For reference, a fully
crystallized sample was made by quenching to 85~K and reheating to
170~K, and a spectrum was measured at all temperatures included in the
study.

\section{Experimental results and data analysis}\label{sec:results}

\subsection{Phenomenological fits of the spectra}\label{sec:fits}

\begin{figure}[ht!]
\centering
\includegraphics[scale=.9]{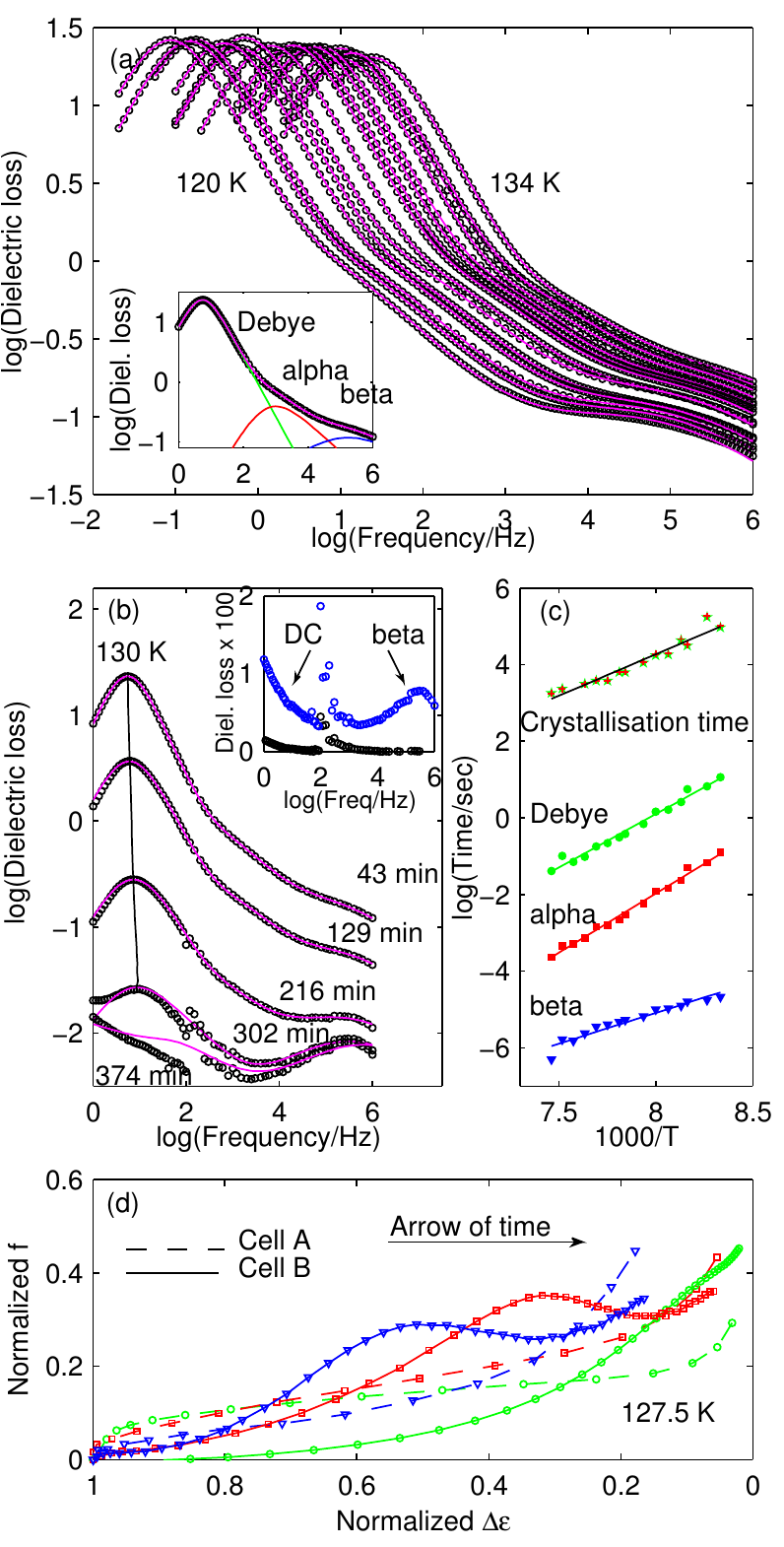}
\caption{\label{fig1}Fitted spectra and time and temperature
  dependence of the fitting parameters. (a) The first measurement at
  every temperature, presumably before crystallization initiates. The
  magenta lines are fits described in the text. The inset show the
  three relaxation processes constituting the fit: the Debye (green),
  alpha (red) and beta (blue) processes.  (b) Selected spectra during
  crystallization at 130 K, using cell B. Again, magenta lines are
  fits. The inset show the dielectric loss of the last measurement
  (blue), together with the measurement of the full crystal (black),
  on a linear scale. The 'spike' at 100~Hz is an experimental artifact
  deriving from a non-perfect match between two voltmeters. (c) The
  fitted relaxation times together with the crystallization time. The
  lines are linear fits. (d) Parameterised plot of the fitted
  relaxation relaxation frequency, $f=1/\tau$, against the fitted
  relaxation strength $\Delta \epsilon$, for each of the three
  processes. The two cells display quite different behavior during
  crystallization.}
\end{figure}

In n-butanol there are three visible processes -- Debye, alpha, and
beta process -- in the measured frequency window, see Fig.\
\ref{fig1}(a). The crystallization, signaled by a decrease of
relaxation strength, also induces a shift in the loss peak for the
three processes. To quantify how the crystallization influences each
of these processes, we fitted the spectra to a sum of three relaxation
processes. Since the processes are not well separated, we aimed at
limiting the number of free fitting parameters by the following
procedure: First the Debye process is fitted by a Cole-Cole
function. The Debye process broadens during the crystallization and
thus a pure exponential function would not give a good fit. The result
of the Debye fit is then subtracted from the data and the alpha and
beta processes are fitted simultaneously as a sum. The beta process is
fitted to a Cole-Cole function with a fixed shape parameter, $\beta =
0.45$. The alpha process is fitted to a dielectric version of the
Extended Bell (EB) model (see Ref. \onlinecite{Saglanmak2010}), in
which the imaginary part of the dielectric constant is given by:
\begin{equation}
  \epsilon''_{\text{EB}} = \dfrac{\Delta \epsilon}{1+ 
    \dfrac{1}{\left(1+i\omega \tau_\alpha
      \right)^{-1}+k_\alpha\left(i\omega \tau_\alpha
      \right)^{-\alpha}}}	 
\end{equation}
where $\Delta \epsilon$ is the relaxation strength and $\tau_\alpha$
is the relaxation time, $k_\alpha$ controls the width of the peak, and
$\alpha$ gives the high-frequency power law behavior of the alpha
peak. This model for the alpha relaxation gives good fits even when
the shape parameters are fixed such that only the relaxation strength
and relaxation time is fitted. The slope parameter was fixed to
$\alpha=0.5$ \cite{Nielsen2009} and $k_\alpha=1$. The Cole-Davidson
function resulted in poorer fits, even with the shape parameter
varying freely. Thus, we fitted the imaginary part to a total of seven
parameters: a relaxation time, $\tau_x$, and a relaxation strength,
$\Delta \epsilon_x$, for each process ($x$ denoting Debye, alpha,
or beta), and a broadening parameter for the Debye process.

The procedure gives excellent fits of over the frequency range
explored as shown in Fig.\ \ref{fig1}(a), where the first dielectric
spectrum at every annealing temperature is shown as well as the fits
resulting from the described fitting procedure. The inset shows a
spectrum at 130~K with each of the individual fitted relaxation
processes.

The same fitting procedure was applied to the isothermal
crystallization spectra to study the temporal evolution of the three
processes during crystallization. This is shown in Fig.\
\ref{fig1}(b) at 130~K using cell B. As the crystallization proceeds
the strength of the Debye and alpha relaxation processes decrease
continuously to disappear entirely by the termination of the
crystallization process. This is also to be expected, since there
should be no large-scale rearrangement of the molecules in the
crystal. The beta relaxation process, however, remains active by the
termination of the crystallization process. The inset of Fig.\
\ref{fig1}(b) shows the last scan at 130~K together with a
measurement of the full crystal at the same temperature, clearly
demonstrating that the crystallization process stops before the sample
is fully crystallized and that there is still some molecular mobility
left. The results from the fitting routine establish the general
behavior of the spectra during the crystallization, but we refrain
from analyzing the finer details, especially towards the end of the
crystallization process, where Debye and alpha process has vanished
and the fits become unreliable (the lowest curve in Fig.\
\ref{fig1}(b)).

Using the relaxation strength, $\Delta\epsilon$, as an indicator of
the degree of crystallinity, we define a characteristic
crystallization time as the time the for $\Delta\epsilon$ to decay to
half of its initial value. Figure \ref{fig1}(c) shows the
crystallization time derived from both Debye and alpha relaxation
strength as a function of inverse temperature, and it is evident that
the two measures are identical within the accuracy of our
measurements.

Along with the crystallization time, we show the relaxation times
obtained from the fits to the first (uncrystallized)
spectrum. Clearly, the Debye, alpha, and beta processes as well as the
crystallization process are all slowed down with decreasing
temperature and consequently the different characteristic time scales
would all appear to be correlated (at least in this temperature
range), but it does not necessarily imply any causation. For the
studied temperature span, all the shown time scales are Arrhenius
within the noise, although with very different pre-factors.


\begin{figure*}
  \centering
  \includegraphics{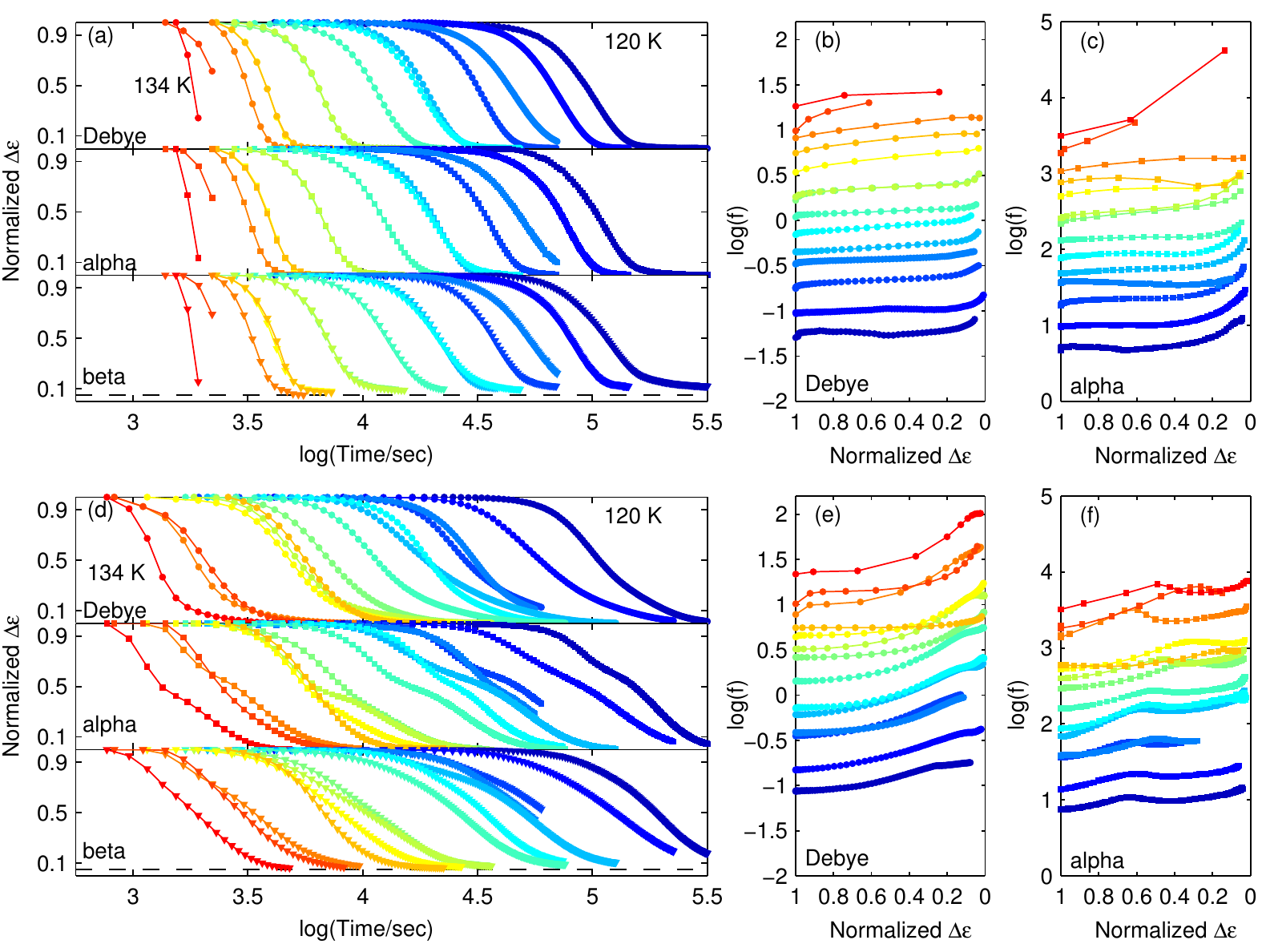}
  \caption{\label{fig3}The fitted relaxation strengths and relaxation
    times for isothermal crystallization curves og n-butanol for
    temperatures between 120 K and 134 K (see text for details on the
    fitting procedure).  The colours of the curves indicate the
    temperatures with blue being the lowest temperature and red the
    highest. 133 K has only been done with cell A while 132 K has only
    been done with cell B. Notice that the crystallization process
    takes longer to finish in cell B.(a+d) The normalized relaxation
    strengths defined as $\Delta \epsilon(t)/\Delta \epsilon(0)$, of
    the Debye, alpha, and beta processes for cell A (a) and B (d) as a
    function of logarithmic time. (b+c+e+f) The logarithm of the
    normalized relaxation frequency ($\log\left[ f(t)/f(0) \right]$,
    with $f = 1/\tau$ ) as a function of the normalized relaxation
    strength for cell A (b+c) and cell B (e+f).}
\end{figure*}

Focusing now on how the two sample environments influence the
crystallization process, we define the relaxation frequency as the
inverse of the fitted relaxation times, $f = 1/\tau$. The evolution of
both the relaxation strength and relaxation frequency differ for the
two cells. Fig.\ \ref{fig1}(d) shows a parameterised plot of the
fitted relaxation frequency (normalized to the initial value) as a
function of the fitted normalized relaxation strength for each of the
three processes from measurements at 127.5~K with both cell A and
B. In cell A, there is a shift in the relaxation frequency, $f$, quite
early in the crystallization process, then it remains relatively
unchanged for all three relaxation processes until an increase sets in
again towards the end of the crystallization. In cell B, the shift in
relaxation frequency is more gradual. For the Debye process the shift
is monotonous, but for the alpha and the beta process the shift
displays a non-monotonous behavior. The curve peaks in Fig.\
\ref{fig1}(d) occurs roughly the same waiting time for the alpha and
beta process. 

The different progress of the crystallization process for the two
cells suggests a macroscopic/mesoscopic rather than microscopic
explanation since a slight difference in sample geometries is not
expected to affect the behavior of individual molecules.

The full set of fitted parameters normalized to the initial value is
shown as a function of waiting time in Fig.\ \ref{fig3}. The colors
of the curves indicate the temperature with blue being the lowest
(120~K) and red being the highest (134~K). In both cells, lower
temperatures lead to longer crystallization times, as was shown in
Fig.\ \ref{fig1}(c).

For cell A (Fig.\ \ref{fig3}(a)) the fitted relaxation strengths for
each of the three processes appear similar, except at long waiting
times, where the beta relaxation strength levels off at $\sim0.08$
instead of decaying all the way to zero. The final level for the beta
relaxation is marked by a dashed line in the third panel of Fig.\
\ref{fig3}(a). For cell B (Fig.\ \ref{fig3}(d)), the Debye and alpha
relaxation strengths follow each other until roughly halfway through
the crystallization, where a shoulder emerges in the alpha relaxation
strength curve, which then proceeds like a two-step relaxation. As in
cell A, the beta relaxation strength does not decay to zero and levels
off at the same value as for cell A. Comparing Fig.\ \ref{fig3}(a) and
(d), we see that the curves for cell B are significantly more
stretched than the corresponding curves for cell B, which means that
crystallization proceeds at a consistently slower rate in cell B
compared to cell A. Consequently, our definition of crystallization
time may give roughly the same for the two cells, however total
crystallization time is much longer in cell B.

Figure \ref{fig3}(b+c) and (e+f) show the parameterized plot
introduced in Fig.\ \ref{fig1}(d) for all temperatures, i.e.,
normalized relaxation frequencies as a function of relaxation
strength, but now separating the two cells and the Debye and alpha
process. (The relaxation frequency of the beta process is not shown,
because it does not vary in a systematic way, making further
interpretation unjustified.)

In cell A, the general behavior is that the relaxation frequencies
have a slight shift to higher frequencies at the onset of the
crystallization, but only increase a little during the remainder of
the crystallization. In cell B, the relaxation frequency of Debye and
alpha processes do not change in the beginning of the crystallization
process, but shifts gradually to higher frequencies. A 'bump' occurs
in alpha relaxation strength around a normalized relaxation strength
of $0.5$, showing that the behavior observed in Fig.\ \ref{fig1}(d) is
general.

The differences between the two cells are thus reproduced for all the
studied temperatures. One mesoscopic explanation for the observed
difference in the evolution of the relaxation strength and relaxation
times for the two cells could be that the two cells induce different
kinds of crystal growth.

\subsection{Maxwell-Wagner analysis}\label{sec:maxwell}

For heterogeneous material, a difference in the conductivity of the
different domains in the material leads to build-up of charges at the
interfaces between domains. This gives rise to a polarization effect
known as Maxwell-Wagner (MW) polarization \cite{Schueller1995, Kremer2003a}.

In the present case the heterogeneity is caused by the formation of
crystallites in the sample. As domains of crystal grow in the liquid
the dielectric constant for the composite will change. The details of
the change will depend on the difference between the dielectric
constant of the liquid, $\epsilon_l$, and that of the crystal,
$\epsilon_c$, the shape of the crystal domain, and the volume fraction
taken up by the crystal.

The two simplest cases of crystal domains growing in the liquid are
that of a crystal layer growing from one (or both) of the electrodes,
a heterogeneous nucleation picture, and the case of crystal spheres in
a liquid matrix, a homogeneous nucleation picture.

In the first case, no approximation is involved in deriving the
expression for the composite dielectric constant. The two materials
(liquid and crystal) in a layered construction are simply modeled by
two capacitors connected in series. Thus the resulting composite
dielectric constant is given by
\begin{equation}\label{model2}
  \epsilon_\text{comp} = \frac{\epsilon_c
    \epsilon_l}{(1-\phi_\text{slab})\epsilon_c +
    \phi_\text{slab}\epsilon_l}\,,
\end{equation}
where $\phi_{\text{slab}}=d_c/d$ is the relative thickness of the
crystal layer. Since $\epsilon_l$ and $\epsilon_c$ can be measured
independently this model has one free parameter (assuming the distance
between the electrodes is fixed, or equivalently that total thickness
of crystal and liquid layer is unchanged during crystallization).

Inserting the measured spectrum of $\epsilon_l$ (at time $t=0$ before
crystallization initiates) and $\epsilon_c$ (spectrum of the fully
crystallized sample), this model produces a large shift in the peak
frequency for even small values of $\phi_{\text{slab}}$. Thus, the
model is unable to capture both the observed decrease in relaxation
strength and the shift in peak frequency at the same time, and this
scenario alone is not sufficient to explain what we observe.

In the case of crystal domains dispersed in a liquid a mean-field
approximation is used to arrive at the composite dielectric constant
\cite{Wagner1914}
\begin{equation}\label{model3}
  \epsilon_\text{comp} = \epsilon_l \frac{2\epsilon_l + \epsilon_c -
    2\phi(\epsilon_l - \epsilon_c)}{2\epsilon_l + \epsilon_c +
    \phi(\epsilon_l - \epsilon_c)} \,,
\end{equation}
where $\phi$ is the concentration of the crystal domains. This model
also contains a single fitting parameter, $\phi$. The mean field
approximation is only accurate up to $\phi \approx 0.2$
\citep{Kremer2003a}, but in the following we allow $\phi$ to go all
the way up to 1.

Inserting the measured spectra of $\epsilon_c$ and $\epsilon_l$ in
Eq.\ \ref{model3} does not produce a frequency shift. Consequently,
this model cannot account for what we observe either.

\begin{figure}[ht!]
  \begin{center}
    \includegraphics[width=6.5cm]{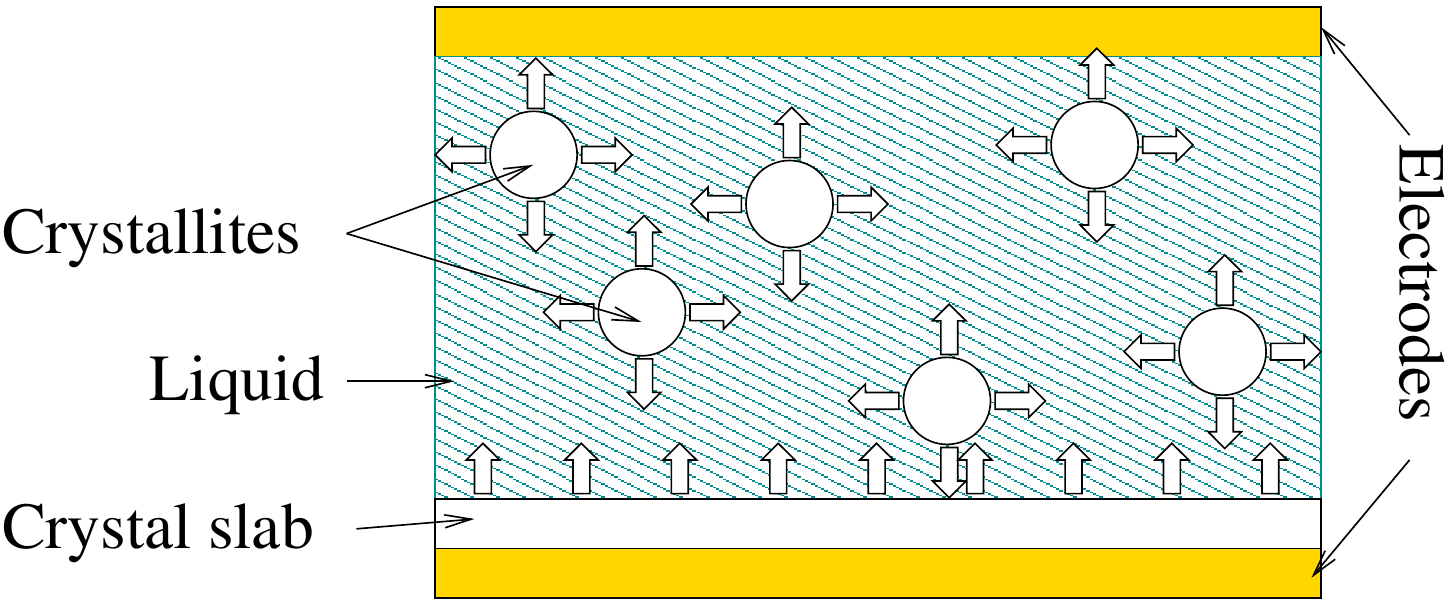}
  \end{center}
  \caption{\label{fig:mw_ill}Cartoon of the combined heterogeneous and
    homogeneous crystallization. The growth is indicated by the
    arrows.}
\end{figure}

Instead, we propose to combine the two models such that a crystal
layer is growing from the electrodes, while spherical crystallites are
forming in the remaining liquid, see Fig.\ \ref{fig:mw_ill}. This is
modeled by combining Eq.\ (\ref{model2}) and Eq.\ (\ref{model3}) such
that $\epsilon_l$ in Eq.\ (\ref{model2}) is given by the composite
dielectric constant from Eq.\ (\ref{model3}). This model has two
parameters: the relative thickness of the crystal layer,
$\phi_{\text{slab}}$, and the concentration of crystal spheres in the
liquid, $\phi$.

\begin{figure}
  \begin{center}
    \includegraphics[width=8cm]{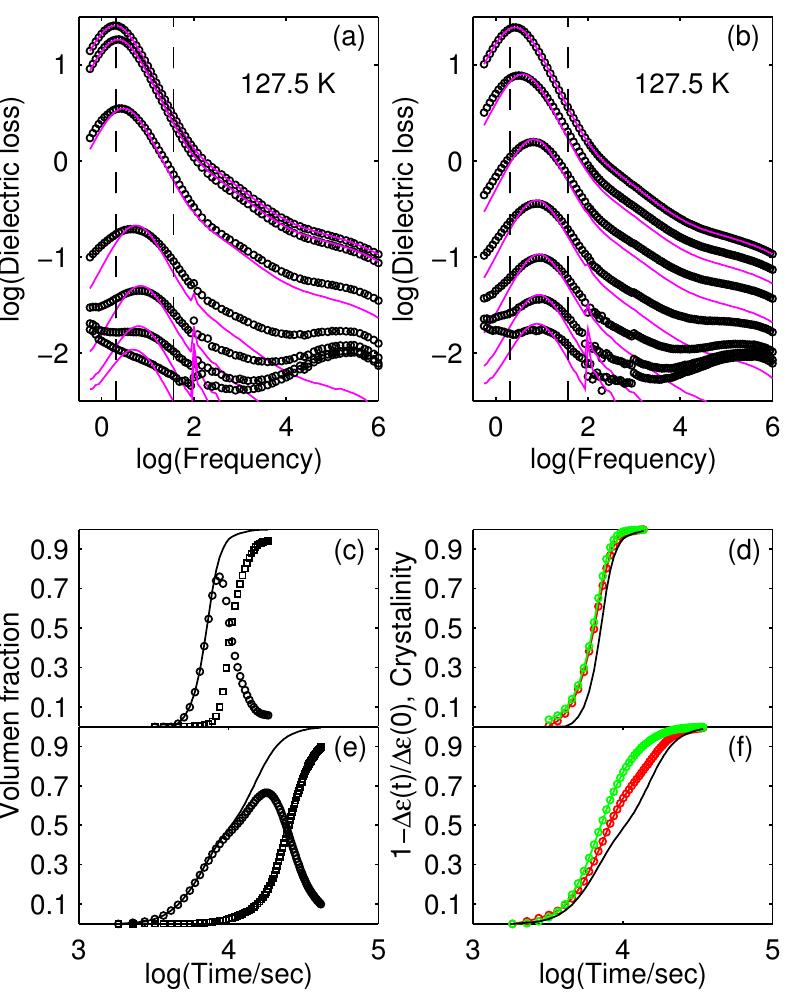}
  \end{center}
  \caption{The measurements shown in Fig.\ \ref{fig1} analyzed using
    the Maxwell-Wagner (MW) approach described in section
    \ref{sec:maxwell}. (a+b) show a selection of curves together with
    fits (magenta lines) for cell A (a) and cell B (b). The dashed
    vertical lines mark the frequency interval used for the fit. (c+e)
    show the volume fraction taken up by the spherical crystallites
    (circles) and the crystal slab (squares) as a function of time for
    cell A (c) and cell B (e). The solid line is the total crystal
    volume fraction. (d+f) Total crystal volume fraction from the MW
    fit (black line) as well as the normalized relaxation strength of
    the Debye process (in green) and alpha process (in red) as a
    function of time for cell A (d) and cell B
    (f). \label{fig:maxwell}}
\end{figure}

Examples of fits to isothermal crystallization spectra for cell A and
cell B are shown in Fig.\ \ref{fig:maxwell}(a) and (b),
respectively. The fits are focused on the Debye process by only
fitting to the points within the two dotted lines. Having two fitting
parameters gives sufficient flexibility to account for both the
decrease in relaxation strength as well as the change in peak
frequency of the Debye peak. However, the combined model does not
adequately explain the behaviour of the entire spectrum; it does not
capture the behaviour of the alpha and beta relaxation processes
during the crystallization or the broadening of the Debye
process. Despite these limitations, the models ability to describe the
behaviour of the Debye relaxation strength and peak position may be
used to shed some light on the observed phenomena.

The total crystallized volume fraction according to the model can be
calculated as $X_c = (1-\phi_{\text{slab}})\phi + \phi_{\text{slab}}$
and the volume fraction taken up by the spheres alone as
$X_{\text{sphere}} = (1-\phi_{\text{slab}})\phi$. Figure
\ref{fig:maxwell}(c) and (e) shows the volume fraction of the crystal
layer and the spherical crystallites as well as the total crystal
volume fraction. For both cells it seems that the onset of
crystallization is dominated by nucleation and growth of crystal
spheres, and when a large fraction of the sample has crystallized
($X_c\sim$80-90\%), the slab growth takes over. However, the growth of
crystal spheres starts earlier but proceeds at a slower rate in cell B
compared to cell A. This difference in crystallization behavior in the
two cells that is reproduced at all temperatures (see Fig.\
\ref{fig:avrami}).

The total degree of crystallinity as calculated from the MW fits is
plotted together with the normalized Debye and alpha relaxation
strengths for cell A in Fig.\ \ref{fig:maxwell} (d) and for cell B in
Fig.\ \ref{fig:maxwell}(f). Again, the curves are clearly different
for the two cells; in cell A Debye and alpha relaxation strengths give
almost identical curves that agree qualitatively with crystal fraction
obtained in the MW fit, although the relaxation strength decreases
faster than crystal fraction initially. In cell B, all three curves
starts out in the same way but separates later in the process, where
both Debye and alpha relaxation strengths overestimates the degree of
crystallinity. The MW crystallinity curve has a kink occurring
approximately when there is a bump in the alpha relaxation strength,
while this two-step behavior is not clearly seen in the Debye
relaxation strength. The behavior of crystallization process at
127.5~K and the MW analysis demonstrated in Fig.\ \ref{fig:maxwell} is
general for all the studied temperatures as can be seen in Fig.\
\ref{fig:avrami}.

The proposed MW analysis qualitatively and quantitatively
agrees with using the decrease of the alpha relaxation strength rather
than the Debye relaxation strength as a measure of the crystallinity
of the sample. However, neither reflect the fact that sample does not
crystallize fully. This is because both the Debye and the alpha
process vanish during crystallization, while only the beta process
survives. The analysis does not account for that, a fact that is
already clear from the fits in Fig.\ \ref{fig:maxwell}(a+b).

\subsection{Avrami analysis}\label{sec:avrami}

\begin{figure*}[ht!]
  \begin{center}
    \includegraphics[width=16cm]{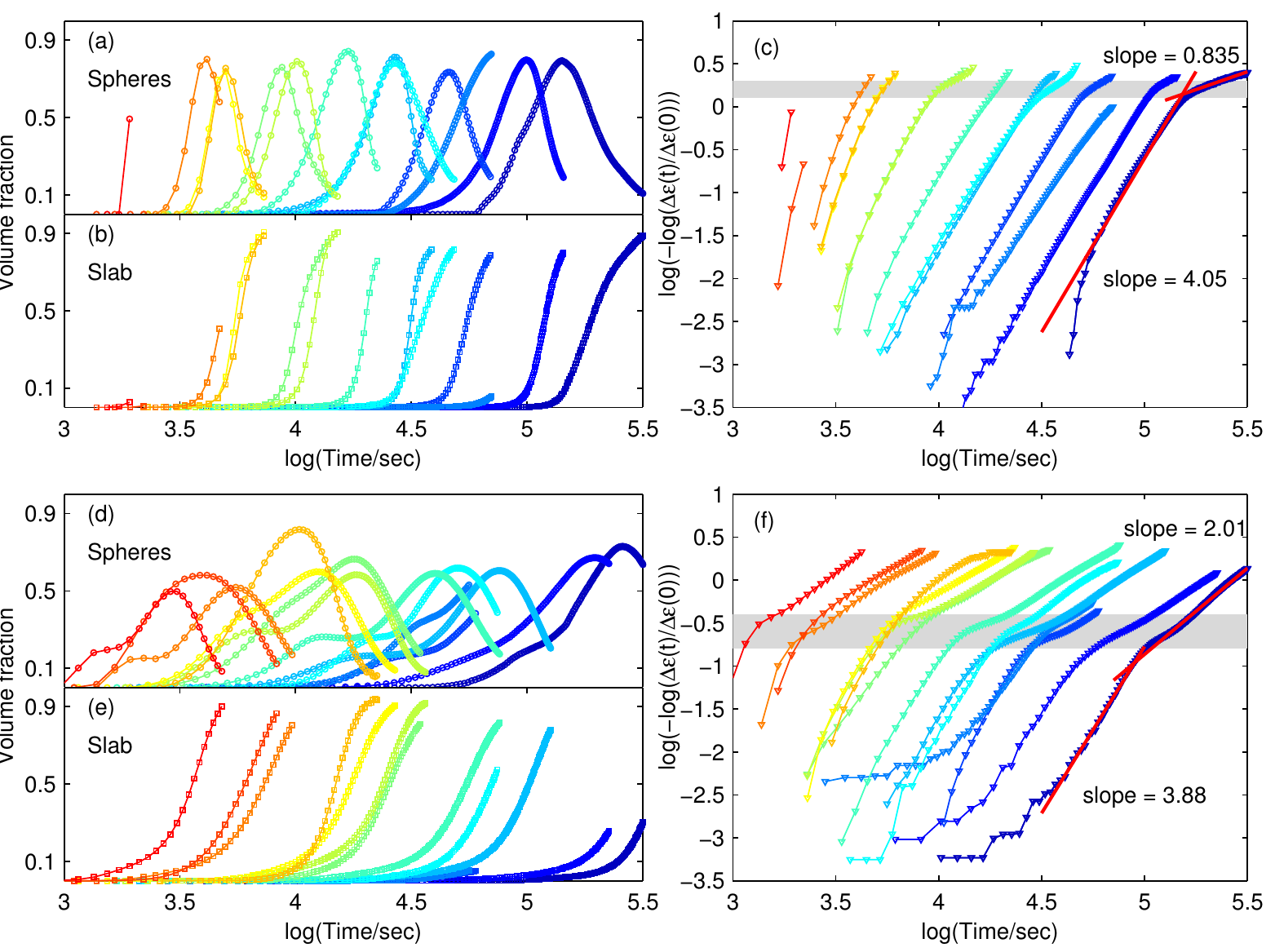}
  \end{center}
  \caption{\label{fig:avrami}Results from MW and Avrami analyses for
    all temperatures. (a+b+d+e) Volume fraction of crystal spheres
    (a+d) and volume fraction of a slab (b+d) based on the proposed
    combination of Eqs.\ (\ref{model2}) and (\ref{model3}). The
    differences for the two measuring cells outlined in Fig.\
    \ref{fig:maxwell}(c+e) are general; cell A (a+b) display a single
    peak in the ``sphere growth'' and the ``slab growth'' set in
    early, while in cell B (d+e) there is a clear double peak in the
    ``sphere growth''. (c+f) show the Avrami-Mehl-Johnson plot which
    is a linearisation of the JMAK equation (Eq.\ (\ref{eq:avrami}) in
    the text). For both cell A (c) and B (f), the data clearly show a
    cross-over from one linear behavior with a steep slope ($\sim 4$)
    at short times to another linear behavior with a smaller slope
    ($\sim$ 1-2). But the transition from one to the other occurs at
    different crystallization degree for the two cells: for cell A the
    transition happens when the alpha relaxation strength has decayed
    more than $95$\% and for cell B when $\Delta \epsilon(t)/\Delta
    \epsilon (0) \sim$40-50\% (marked by the gray bars in the
    figure).}
\end{figure*} 

Another independent -- and more routinely used -- way of evaluating
the crystal growth is through the Johnson-Mehl-Avrami-Kolmogorov
(JMAK) equation originally developed by Avrami \citep{Avrami:1939cn,
  Avrami:1940hb, Avrami:1941ff}. In this analysis, the volume fraction
taken up by crystallites, $X_c$ is expressed in terms of a growth rate
constant $k$, an induction time $t_0$, and the Avrami parameter $n$ as
follows
\begin{equation}\label{eq:avrami}
	X_c(t) = 1 - \exp\left[ -k\left(t-t_0 \right)^n \right]\,.
\end{equation}
The value of $n$ depends on the crystal morphology and crystallization
mechanism, but it is not straight forward to interpret the meaning of
this parameter. Originally, it was a number between 1 and 4 such that
$n = d+r$ with $d$ being the dimensionality of the growth and $r$
being a number that represents the nucleation rate. For a constant
nucleation rate $r=1$ and $r=0$ if nucleation stops when the
crystallization starts. More recently $n$ has been found to be a
number between 1 and 7 \citep{Viciosa:2009hr, Dantuluri:2011gu}.

When dielectric spectroscopy is used to study crystallization the
common practice is to assume that the alpha relaxation strength
roughly corresponds to the degree of crystallinity, $X_c$, and then
use that for the Avrami analysis. Since the MW model also gives some
support for the alpha rather than the Debye relaxation strength is
expressing the degree of crystallinity in the sample, we will adopt
this approach.

One way to obtain an estimate of the parameter $n$ is through the
Avrami–Mehl–Johnson plot, which plots $\ln \left[ -\ln( 1-X_c(t))
\right]$ versus $\ln(t)$. This procedure avoids fitting Eq.\
(\ref{eq:avrami}) to data and $n$ is directly obtained as the slope of
the curve. The Avrami-Mehl-Johnson plots for cell A and B is shown in
Fig.\ \ref{fig:avrami}. In both cases, we see a transition from a
relatively high value $n\approx 4-5$ to a low value $n\approx
1-2$. This observations suggests a change from higher dimensionality
of growth to lower, which is consistent with the MW analysis
suggesting a change from spherical to slab growth. Moreover, the data
suggest that this transition happens earlier in the crystallization
process for cell B compared to cell A, which could explain why the
crystallization slows down and takes much longer in cell B.

\section{Discussion}\label{sec:disc}

Both the Debye process and the alpha process vanish during the
crystallization, while the beta process survives. Thus we confirm
earlier findings that the crystallization process at temperatures near
$T_g$ stops before the sample is completely crystallized. Hedoux
\textit{et al} report signs of an aborted or frustrated
crystallization process, signaled by a amorphous halo persisting in
the x-ray spectra \cite{Hedoux:2013ep}. This slow and frustrated
crystallization process has also been interpreted as a polyamorph
transformation between two meta-stable liquid phases
\citep{Dzhonson:2003tp, Bolshakov:2005ja, Kurita:2005jr,
  Zgardzinska2010}. Based on the dielectric spectra presented here it
is perhaps difficult to distinguish between the two scenarios, but the
fact that the structural relaxation peaks disappear entirely combined
with the emergence of Bragg peaks as documented in Ref.\
\cite{Hedoux:2013ep}, point to a non-trivial crystallization process
as the most obvious explanation for the observations.

It is however interesting that the aborted crystallization is seen in
the dielectric spectra as the survival of the beta-process. If we
envision the end product as a frustrated crystal, unable to tile
space, then the liquid signal -- in our case the beta relaxation --
could originate from small pockets of liquids between crystal
grains. This picture supports the idea of the beta relaxation being a
local phenomenon, in favor of the ``islands of mobility'' suggested by
Johari and co-workers \cite{Johari1970, Johari2002}.

The mono-hydroxyl alcohols in general are interesting because of their
anomalous (and usually intense) relaxation process at frequencies
lower than the structural alpha relaxation -- the so called Debye
relaxation -- which is believed to be due to supra molecular hydrogen
bonded structures in the liquid \cite{Gainaru2014, Bohmer2014}. We
observe that the Debye process vanishes faster than the alpha during
crystallization, and that the alpha intensity seems to give a better
measure for the degree of crystallinity. Sanz \text{et al}
\cite{Sanz2004} made similar observations for another monohydroxyl
alcohol. They studied crystallization of isopropanol in real time by
simultaneous dielectric spectroscopy and neutron diffraction
measurements, and thus had a direct measure of the degree of
crystallinity that could be correlated with the relaxation strength of
the Debye and alpha process. They observed that the Debye intensity
dropped rapidly at the onset of crystallization, while the alpha
intensity followed the crystallization. Their intuitive and appealing
interpretation was, that the breakage of the hydrogen-bonded network
is a precursor of the crystallization, and that the molecules leaving
the network did not immediately go into a crystalline structure. The
MW polarization effects lends itself to a different -- macroscopic --
interpretation of the observations. Irrespective of how the crystal
growth is modeled in the MW framework, there cannot be proportionality
between dielectric intensity and liquid fraction in the sample. The
deviation from linearity depends on the specific model for the growth
morphology and on the intensity of the process: the higher intensity,
the stronger the deviation from linearity. Thus the MW analysis
provides a simple explanation for why the most intense process vanish
before the less intense one. The MW analysis does at the same account
for the observed frequency shifts of the relaxation processes.

On the basis of MW fits, we suggested that the observed difference in
crystallization behavior between the two cells could be rationalized
by a transition from having primarily a growth of crystal spheres, a
homogeneous nucleation and growth, to a growth of a crystal slab. This
could be a slab growing from the electrodes, but could also be a
certain point in the process where crystal grains percolate and
effectively create a crystal layer in the liquid-crystal mixture. The
difference between the cells would then be explained by a difference
in the degree of crystallization when this transition takes place. The
idea that the crystal growth changes from a higher dimensional growth
to a low dimensional growth was supported by the Avrami–Mehl–Johnson
analysis that also points to such a transition taking place at
different crystallization degrees in the two sample cells. The overall
validity of JMAK equation has been questioned, see e.g. Refs.\
\onlinecite{Todinov:2000uh, Fanfoni:1998wu}, and of course we need to
be cautious when making conclusions, based on the MW analysis where
the limits of applicability of the mean-field approximations was
pushed. But since both types of analyses point with this picture, we
believe that the proposed conception of a change in morphology of the
crystal growth is consistent and sound. It remains to be shown how
general this behavior is. It would be interesting to apply this
procedure to a simpler sample to study the influence of sample cell
geometry on the crystallization process. 

Irrespective of the generality of the particular behavior found here,
our study shows that one needs to be very cautious about making
detailed microscopic interpretations of the crystallization mechanisms
based on dielectric spectroscopy alone, because MW polarization
effects of the mixed phase requires knowledge about the crystal growth
morphology. In addition, we have also shown that the crystallization
is extremely sensitive to the specific sample environment. Thus, it
would require extensive investigations of different environments and
perhaps even different probes to disentangle microscopic from
macroscopic effects.

\section{Conclusion}

We have studied isothermal crystallization process in the deeply
supercooled region of the mono-alcohol n-butanol in real time at 15
different temperatures using dielectric spectroscopy. Two different
sample cells have been used to look for the effects of the sample
environment on the crystallization process. We found that the time
evolution of the relaxation strengths differs for the two cells in a
consistent and reproducible way for all temperatures.

On the basis of the Maxwell-Wagner analysis, we suggest that the
crystallization behavior can be explained by a transition from
primarily growth of crystal spheres to growth of a crystal layer. The
difference between the cells in this framework is the difference in
when in the crystallization process this transition takes place. This
picture was supported by an Avrami–Mehl–Johnson analysis that also
suggests a transition from higer dimensional growth to a lower one.

The Maxwell-Wagner analysis can also account for the shift in peak
frequency observed for the three processes during the course of
crystallization, and thus a microscopic interpretation of the peak
shift is not needed.

\section{Acknowledgements}
Tina Hecksher is sponsored by DNRF Grant no 61.  The authors thank
Ranko Richert for useful discussions and valuable comments.

%


\end{document}